# Instantaneous frequency estimation in compressive sensing scenario

(Student paper)

Božidar Andrović, Marko Kovač and Anđela Kandić
Faculty of Electrical Engineering
University of Montenegro
Podgorica, Montenegro

*Abstract*—In this paper, the instantaneous frequency estimation of nonstationary signals is considered. The instantaneous frequency is estimated from the time-frequency representation where certain percent of the coefficients is missing. The time-frequency representation is considered as an image, whose missing pixels are reconstructed by using compressive sensing recovery algorithms. As a time-frequency representation, the S-method is considered. The Compressive Sensing as a intensively growing novel approach for signal acquisition, ensures accurate signal reconstruction from relatively small percent of available information about the signal. The theory is verified by experimental results.

I. INTRODUCTION

Different forms of time-frequency (TF) distributions have been used for the non-stationary signals analysis [1]-[10]. The commonly used TF representation is the short-time Fourier transform (STFT), whose energetic version is called spectrogram. The spectrogram is relatively simple for realization, but it has some drawbacks such as window width dependent resolution in the TF plane. Hence, the quadratic TF distributions have been introduced in order to overcome this problem and to provide better concentration of the instantaneous frequency (IF) in the TF domain [6], [8]. The commonly used quadratic distribution is the Wigner distribution. It provides an ideal representation for linear frequency modulated signals. However, when it comes to multicomponent signals, it suffers from undesired components called the cross-terms. Also, in the cases of fast variations of the IF, the Wigner distribution can introduce an inner interferences, that are also the unwanted terms. Therefore, the S-method has been introduced to overcome those shortcomings of the Wigner distribution. By using the S-method and by optimal selection of the window width in the S-method, the cross-terms could be reduced and in some cases completely removed. Here, we will focus on one component signals, and multicomponent ones can be subject of our future research.

Due to certain limitations of traditional sampling theory based on the Shannon-Nyquist sampling theory (regarding that the sampling rate must be at least twice the maximum signal frequency), the Compressive Sensing (CS) theory have emerged as a huge improvement [11]-[23]. The main goal of the CS is to enable the signal reconstruction from much fewer number of samples compared to the number of samples required by the Sampling theorem. A lot of work, time and efforts are put into making that possible, and therefore, we may say that the CS was a big breakthrough. Not only for lowering the amount of samples required for signal reconstruction, but it also provides the opportunity to simplify very expensive apparatus and devices for data recording and sensing.

The paper is organized as follows. In Section II describes the time-frequency representations used in the paper. The CS approach is mathematically described in the Section III, along with the reconstruction algorithm used for recovery of TF images. In Section IV, the IF estimation of the original and the reconstructed signal are experimentally verified while the concluding remarks are given in Section V.

II. TIME-FREQUENCY ANALYSIS

The short-time Fourier transform (STFT) is the commonly used time-frequency signal representation. The STFT is obtained by sliding the window $w(t)$ along the analyzed signal $x(t)$. Mathematically this distribution can be described as follows [6], [8]:

$$STFT(t,w) = \int_{-\infty}^{+\infty} x(t+\tau)w(\tau)e^{-jw\tau}d\tau \quad (1)$$

A square module of the STFT is called the spectrogram and it is defined as:

$$SPEC(t,w) = |STFT(t,w)|^2 \quad (2)$$

However, in most cases the spectrogram cannot provide a satisfactory time-frequency resolution. Namely, the resolution depends on the window width. Also, this distribution is not suitable for fast varying signals. Therefore, other distributions are used, such as the Wigner distribution, S-method, higher order distributions etc. Here, we will focus on the S-method (SM) [4], [7], [8], the distribution that can provide good resolution in the time-frequency plane, even for the fast varying signals. The SM is defined as [4], [8]:

$$SM(t,w) = \frac{1}{\pi} \int_{-\infty}^{+\infty} P(\theta) STFT(t, w+\theta) STFT^*(t, w-\theta) d\theta \quad (3)$$

where $P(\theta)$ represents a finite frequency domain window function. The discrete version of the S-method is:

$$SM(n,k) = \sum_{l=-L}^{L} P(l) STFT(n, k+l) STFT^*(n, k-l) \quad (4)$$

where $n$ and $k$ are the discrete time and frequency variables, respectively, while $P(l)$ is the window of the length $2L+1$. By taking the rectangular window, the discrete S-method can be written as:

$$SM(n,k) = |STFT(n,k)|^2 + 2\mathrm{Re}\left\{\sum_{l=1}^{L} STFT(n, k+l) STFT^*(n, k-l)\right\} \quad (5)$$

The window $P(l)$ should be wide enough to enable the complete summation over the auto-terms. At the same time, in order to remove the cross-terms, it should be narrower than the minimal distance between the auto-terms.

### III. COMPRESSIVE SENSING

Compressive Sensing is signal sensing approach that provides signal analysis and reconstruction using small number of randomly chosen samples [10]-[18]. Reconstruction is based on complex mathematical algorithms - optimization algorithms. Depending on the signal type, different optimization algorithms are used [8], [15]-[17], [19].

Finite, real, 1D signal $x$ could be described as column, $N$x1 vector, in the $R^N$ space. Mathematically, signal could be described as:

$$x = \sum_{i=1}^{N} s_i \psi_i = s\psi \quad (6)$$

where $\psi$ denotes $N$x$N$ transform domain matrix, while $s$ is transform coefficients vector. Vectors $x$ and $s$ are representations of the same signal in different domains - in time domain and in transform domain. Two conditions have to be satisfied in order to reconstruct signal from the small number of acquired samples. First, the signal has to be sparse, which means that small number of signal coefficients (in its own domain or in the transform domain) has non-zero values. The second condition is related to the measurement procedure. Namely, the measurement procedure has to be incoherent. If it is satisfied, the signal could be reconstructed with high accuracy using small number of samples.

The number of acquired signal samples (measurements) can be much smaller than the signal length $N$, i.e. $M<<N$. Measurement vector $y$ is obtained by multiplication of the measurement matrix $\phi$ by signal vector $x$, which could be described as:

$$y = \phi x \quad (7)$$

Combining the relations (6) and (7) the following equation is obtained:

$$y = \theta s \quad (8)$$

where $\theta = \psi\phi$ is the CS matrix. System of (8) has infinite number of solutions. In order to obtain a unique solution, the optimization algorithms are used. One of the commonly used is optimization based on the $l_1$-norm minimization [8]:

$$s = \min \|s\|_{l_1} \quad \text{subject to} \quad y = \theta s \quad (9)$$

*Algorithm for the CS image reconstruction*

As it was mentioned before, the signal can be reconstructed from its measurement using complex optimization algorithms. Commonly used optimization, based on $l_1$ minimization, does not provide satisfactory results when applied on image The image is generally not sparse in any transform domain, but the image gradient can be considered as sparse. Therefore, the $l_1$ minimization of the image gradient is performed and it is called the TV minimization [8], [17], [21]-[23].

TV of the signal $s$ is sum of the gradient amplitudes in the point $(i,j)$. It can be described as:

$$\|s\|_{TV} = \sum_{i,j} |(\nabla s)_{i,j}| \quad (10)$$

where $\nabla$ represents differentiation operator, i.e. approximate value of the gradient, for pixel (i,j):

$$\nabla_{i,j} s = \begin{bmatrix} s(i+1, j) - s(i, j) \\ s(i, j+1) - s(i, j) \end{bmatrix} \quad (11)$$

Discrete form of the TV could be described as:

$$TV(s) = \sum_{i,j} \sqrt{(s_{i+1,j} - s_{i,j})^2 + (s_{i,j+1} - s_{i,j})^2} \quad (12)$$

TV minimization problem is given by (13):

$$\min TV(s) \text{ subject to } y = \theta s \quad (13)$$

TV minimization provides a good reconstruction of the signal and gives acceptable results in the cases of noisy signals also.

### IV. EXPERIMENTAL RESULTS

In this assignment, we consider three signals. The first one shows the slow variations of the instantaneous frequency, the second example is the signal with fast instantaneous frequency variations while the third one is the chirp signal. The signals are described by the following relations:

- $x(t) = e^{j\sin(1.2\pi t)}$
- $x_1(t) = e^{j\sin(10\pi t)}$
- $x_2(t) = e^{-j40\pi t^2}$

As a TF representations, the S-method is used. The TF representation obtained with the SM provides better temporal and frequency resolution, which allows us to obtain more accurate description and estimation of IF.

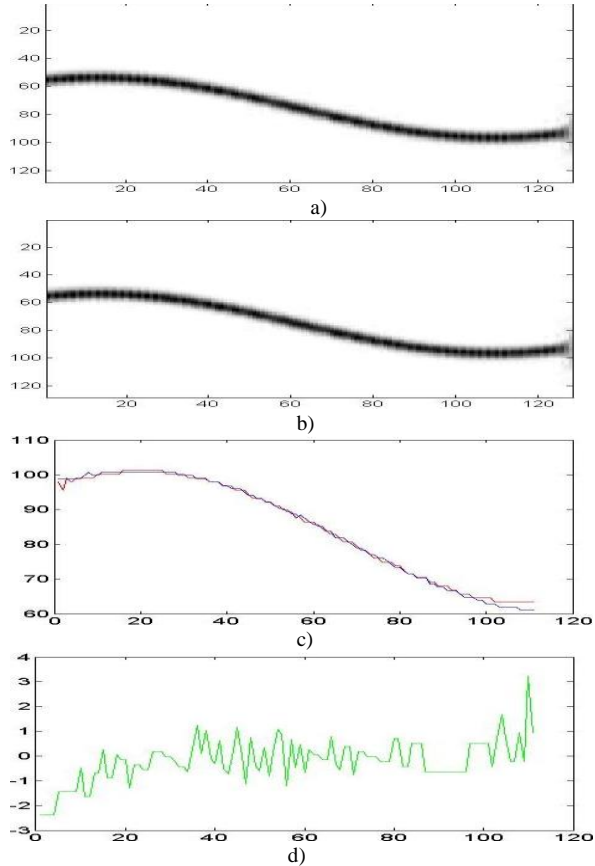

Fig.1 a) S-method, b) the reconstructed S-method using 46% of the total number of samples, c) IF of original signal-red, IF of reconstructed signal-blue line, d) error between original and reconstructed IF

The procedure is applied as follows. First, the S-method matrix is converted into jpeg format. The S-method is therefore, observed as an image. This image is damaged in a way that certain amount of pixels are discarded. The under-sampled image is recovered by using the CS recovery algorithms, i.e. the TV minimization. After the image reconstruction, the IF is estimated and compared to the IF estimated from the original S-method image.

The signal with slow variations of the IF is observed first. The results obtained by using the S-method and the reconstructed S-method are shown in Fig. 1.

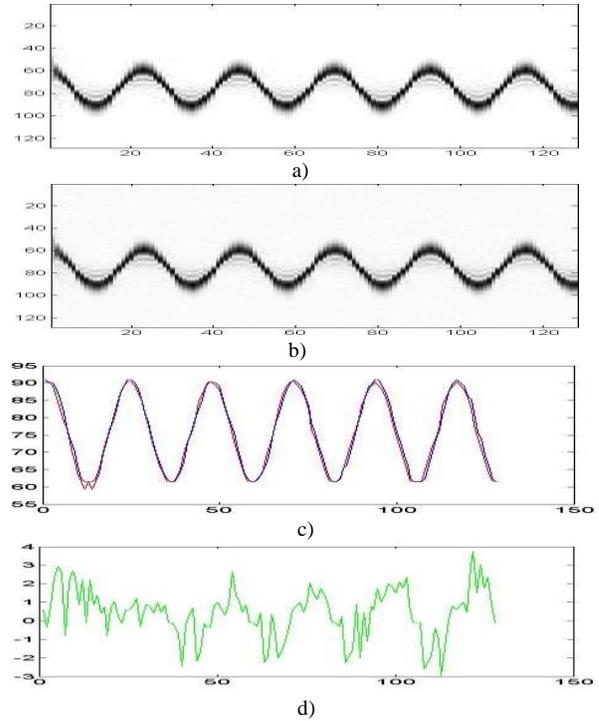

Fig.2 a) The S-method of the original signal, b) The S-method of the reconstructed signal using 46% of the total number of samples, c) IFs of the original (red) and the reconstructed signal (blue), d)error between original and reconstructed Ifs

The S-method is recovered using 45% of the high-frequency and 1% low-frequency samples. The original and IF estimated from the reconstructed image are shown on the same graph in Fig 1c. The difference between two signals is measured and compared. It is in fact error signal, which is shown in green line on each graphic.

Fig. 1 refers to signal $x(t)$. Fig.2 is related to the signal $x_1(t)$ and Figs.3, 4 and 5 are related to the $x_2(t)$.

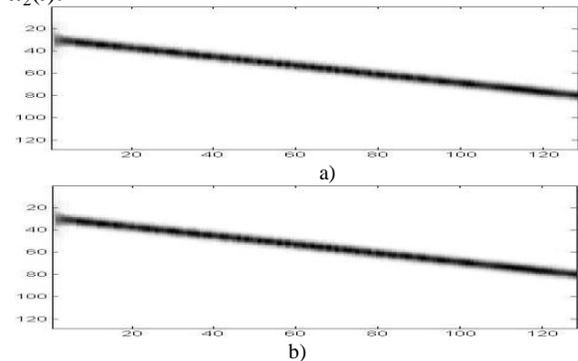

Fig.3 a) The S-method of the original signal, b) the reconstructed S-method using 45% of the high-frequency and 1% low-frequency samples

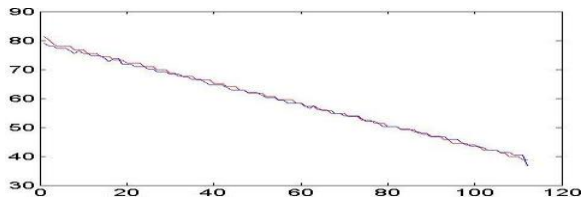

Fig.4 The IF of original signal-red, the IF of reconstructed signal-blue line

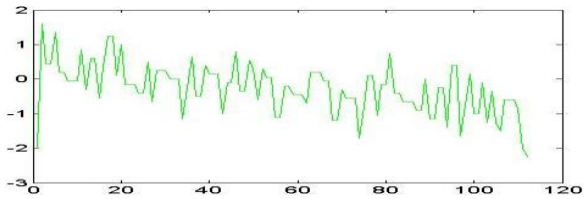

Fig.5 The error between the original and the reconstructed IF

## V. CONCLUSION

Application of CS approach for IF estimation is considered in this paper. Different kinds of signal are observed. The IF estimation of each signal is performed before and after image reconstruction, since the TF matrix is observed as a jpeg image. Namely, the original TF representation is firstly converted to image and then it is under-sampled. The image is recovered using relatively small percent of image pixels. The IF estimation form recovered image is successful and it is approximately the same as the IF of original signals. The future work will be focused on CS application in IF estimation of different types of signals, using other forms of the TF distributions.